\documentclass[twocolumn]{aastex62}
\usepackage{graphicx}

\begin{document}

\title{Serendipitous Detection of HI Absorption Sets the True Redshift of 4C +15.05 to $z=0.833$}

\correspondingauthor{K.~M.~Jones}\email{kjones@naic.edu}
\author{K.~M.~Jones}\affil{Arecibo Observatory, HC-03 Box 53995, Arecibo PR 00612}
\author{T.~Ghosh}\affil{Arecibo Observatory, HC-03 Box 53995, Arecibo PR 00612}
\author{C.~J.~Salter}\affil{Arecibo Observatory, HC-03 Box 53995, Arecibo PR 00612}

\begin{abstract}
4C+15.05, (also known as NRAO 91, PKS 0202+14 or J0204+15), is a quintessential blazar. It has a luminous, variable radio spectrum, a super-luminal jet, and gamma-ray detections. Arecibo observations with the 700-800 MHz receiver on the 305-m diameter William E. Gordon Telescope detected, serendipitously, HI in absorption against 4C+15.05 while using it as a bandpass calibrator for another object in an HI absorption project. Although the redshift we derive is different from that commonly in use in the literature (nominally $z=0.405$), it agrees very well with the value of $z=0.833$ determined by \cite{Stickel+96}. This absorption feature is best fitted by a sum of three Gaussians, which yield an average redshift of $z=0.8336 \pm 0.0004$, although without corresponding high resolution imaging it is not possible to say whether the components are parts of outflows or inflows. A total column density of $N(HI) = 2.39 \pm 0.13 \times 10^{21}$ cm$^{-2}$ is derived, relatively high compared to many radio-loud sources. These results are compared to various relationships in the literature. 

\end{abstract}

\section{Introduction}
The blazar 4C+15.05 has been studied extensively since its inclusion as a bright ($\sim$ 5 Jy at 178 MHz) radio source in the 4C catalog \citep{PT66,Gower67}.  Early radio studies found a relatively flat spectrum, with a 318-5000 MHz slope of $\alpha < +0.5$, where $S_{\nu} \propto \nu^{-\alpha}$, and \cite{Condon+1974a,Condon74b} found $\alpha(318,7200)= +0.11 \pm 0.18$ and $\alpha(318,5000)= +0.26 \pm 0.02$, respectively. The source remains bright even at millimeter wavelengths \citep[][and references therein]{Kellermann+71,Owen+1980}. This flatness, associated with the compact cores of bright radio sources, is attributed to the effects of source non-uniformity within the core, either in the form of distinct physical components, or in the form of distinct relativistic electron or magnetic field distributions in the environment of the AGN \citep[e.g.][]{Kellermann+PT69,Condon+1973,Marscher+1977,Giovanoni+Kazanas90}. 4C+15.05 was found to demonstrate the characteristic short-timescale variability of blazars -- first at low frequencies (e.g., \cite{Ghosh+Rao92} found a variability index of -2.5\% over 15 years at 327 MHz) and then, as data became available, at higher frequencies (with a variability index of 4.2\% over 2 years at 5 GHz \citep{Bennett84};  with a variability index of 6.4\% over 5 years at 22 GHz and 14.8\% over 5 years at 37 GHz \citep{Terasranta+1992,Terasranta+98,Terasranta+2000,Terasranta+2005}). This variability extends into the infrared, e.g., \cite{Stickel+96} classify it as violently variable in the J, H, and K bands, and demonstrates the pervasive influence of the AGN on the broadband spectral emission. This blazar, unsurprisingly, exhibits polarization both in the radio (\cite{Fan+2008} report a polarization of 5.2\% at 8 GHz) and the optical (e.g., \cite{Impey+Tapia90} report a polarization of 3.2\% at $\sim$ 5500 \AA). Super-luminal motion of the interior jet is detected in \cite{Lister+13}, who report speeds of up to 15.88c. 4C+15.05 is also a known $\gamma$-ray source, and ~\cite{Linford+12} report a $\gamma$-ray flux $ = 29.98$ ($\pm 12.42) \times 10^{-9}$ photons cm$^{-2}$ s$^{-1}$.

As with other AGNs, the redshifts of blazars were initially difficult to determine. However,  the extensive use of spectroscopy and the understanding of blazars as extra-galactic objects resolved this issue. Previous work has shown diverse values for the redshift of 4C+15.05, with the earliest estimate being from ~\cite{Dermer+92}, who reported $z=1.202$ (labelled ``uncertain''). \cite{Stickel+96} were the first to report a redshift derived from near-infrared/optical spectroscopy, namely $z=0.833$. This was followed by ~\cite{Perlman98} whose slightly expanded NIR/optical spectra led them to report $z=0.405$. Given the expanded spectrum and improved resolution of the Perlman et al. work, this third value has been adopted by the community at large as the correct value. The NASA/IPAC Extragalactic Database (NED) quotes $z=0.405$ and subsequent papers that employ a redshift for 4C+15.05 have used this lower value.

We now report a new independent measurement that determines the true redshift for 4C+15.05 to be $z=0.8336 \pm 0.0004$.

\section{Observations and Results}
\subsection{Details of the Observations}
In November 2010, taking advantage of the temporary freeing up of the 700-800 MHz band in June 2009 due to the analog-to-digital TV transition, a special 700-800 MHz receiver was installed at Arecibo Observatory. We used that system to target eight strong radio-loud galaxies whose redshifts move the 21-cm HI line into the passband of this receiver. 

The purpose of the project was to search for HI absorption as a signature of gas in the nuclear regions and then follow up those objects with positive detections with higher resolution observations to probe the kinematics of neutral nuclear gas as a means to study the fueling mechanisms of supermassive black holes. The dual-polarization 700-800 MHz receiver was cooled to liquid nitrogen temperatures, yielding a T$_{sys} \sim 110 $ K. It operated from mid-2008 to early 2012; after that date, wireless and emergency communication channels made this frequency regime inaccessible for astronomical observations.


4C+15.05's expected redshift of $z=0.405$ did not move its HI transition into this band; it was thus not a target, but was included as a bandpass calibrator source for the nearby BL Lac object, B0235+164.  The Double Position Switching (DPS) technique \citep{Ghosh+Salter02} was used for each source. This technique is used to remove the baseline ripple created by an object's own radio emission, which creates standing waves within the unique structure of the 305m telescope and its platform. DPS obtains position-switched spectra of a target source and a calibrator source at similar azimuths and zenith angles in order to remove this ripple, allowing the detection of low-level absorption. Each 5-min ON/OFF target observation (5 min on the target source, and 5 min on blank sky) was therefore matched by a 5-min ON/OFF calibrator cycle (5 min on the calibrator source, and 5 min on blank sky); the minimum expected observing time for a single cycle with this technique is therefore about 25 min, allowing for slewing. 4C+15.05 was observed on 20, 21, and 22 November 2010, with clear skies, for a total on-source integration time of approximately 50 minutes. 

\subsection{Results}

\begin{figure}
\begin{center}
\includegraphics[scale=0.5,trim= 0cm 0cm 0cm 1cm ,clip=true]{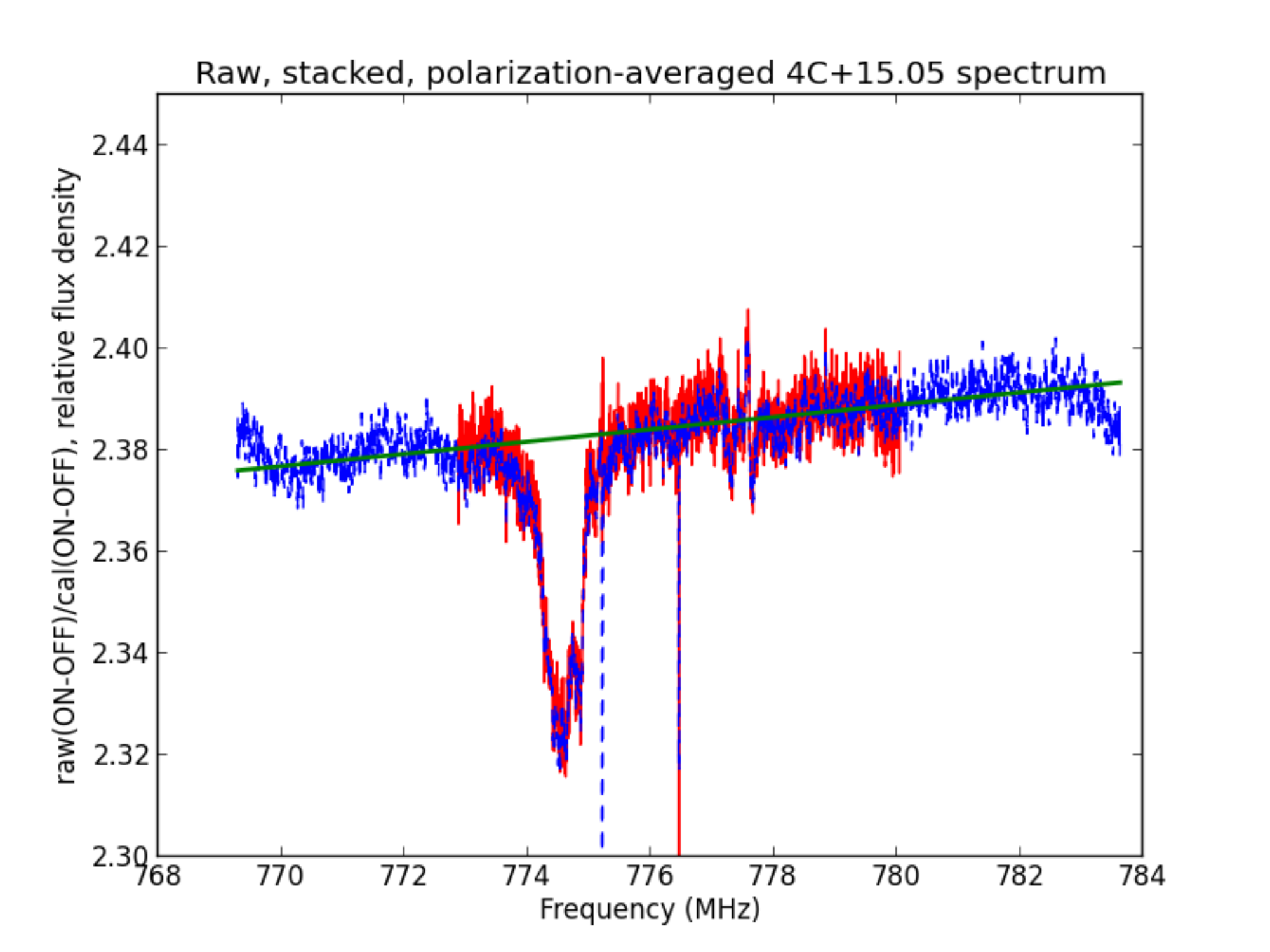}
\caption{\label{fig1} Stacked, polarization averaged raw spectra from the 700-800 MHz receiver. The red data show the highest-resolution spectrum; the blue the second-highest-resolution spectrum; both are smoothed with a boxcar function having a 9-channel width. The green line is the fitted baseline (described in Section 2.2). The spike at the center of the spectrum is an artifact of the spectrometer. The $y$-axis is relative flux density, obtained by averaging the on-off spectra for the target source and for the bandpass calibrator, then dividing these to remove baseline ripples.}
\end{center}
\end{figure}

Individual ON/OFF scans of 4C+15.05 were processed using the Arecibo IDL analysis packages written by Phil Perillat and Robert Minchin to yield bandpass-corrected spectra, using B0235+164, which showed no HI absorption, as the bandpass calibrator.  These spectra were then co-added and the orthogonal polarizations combined to produce the final `raw' spectrum, which is presented in Figure 1; the $y$-axis is a relative flux density produced by $(ON-OFF)_{target}/(ON-OFF)_{cal}$.  Absorption was detected at $z \sim 0.83$ with a fractional absorption of $\sim$ 2.5\%. Using a baseline fitting routine that is part of the Arecibo analysis suite, a first order polynomial (i.e., linear) baseline was fitted to the continuum portion of the spectrum (after masking out obvious RFI and the frequency range of the HI absorption). Third, fourth, and fifth order polynomials were also fitted to the spectrum continuum but were not an improvement. The fitted baseline was then used to provide the unabsorbed continuum level within the frequency range of the HI absorption. This allowed the derivation of optical depth ($\tau_{\nu}$, from $S_{\nu,absorbed}=S_{\nu,continuum}e^{-\tau_{\nu}}$). Integrating $\tau$ over the width of the line in km/s then gives an estimate of the column density of the HI lying in front of the AGN continuum emission:

\begin{equation}
N(HI) = 1.835 \times 10^{18} \frac{T_{s}\int\tau dv}{f_{c}} cm^{-2}
\end{equation}

In addition, the absorption line was well fit by a combination of three Gaussians, as shown in Figure 2. 
Assuming the HI spin temperature $T_{s} = 100$ K and the covering fraction $f_{c} = 1.0$, and integrating over the optical depth spectrum yields a total column density of $N$(HI)$ = 2.39 \pm 0.13 \times 10^{21}$ cm$^{-2}$. The residuals presented in Figure 2 between the derived optical depth and the sum of the three Gaussian fits indicates the quality of the fit. The parameters of the three components are presented in Table 2. The column density for each component has been determined using the approximation that for $\tau (v)$ that takes the shape of a Gaussian, $1.835 \times 10^{20} \int\tau (v) dv$ cm$^{-2}$ $= 1.93 \times 10^{20} \tau_{peak} \Delta v$ cm$^{-2}$, where $\Delta v$ is the FWHM and $\tau_{peak}$ is the peak optical depth of each component (under the continued assumption of $T_{s}$ = 100 K and $f_{c}$ = 1.0).


\begin{figure}
\includegraphics[scale=0.4]{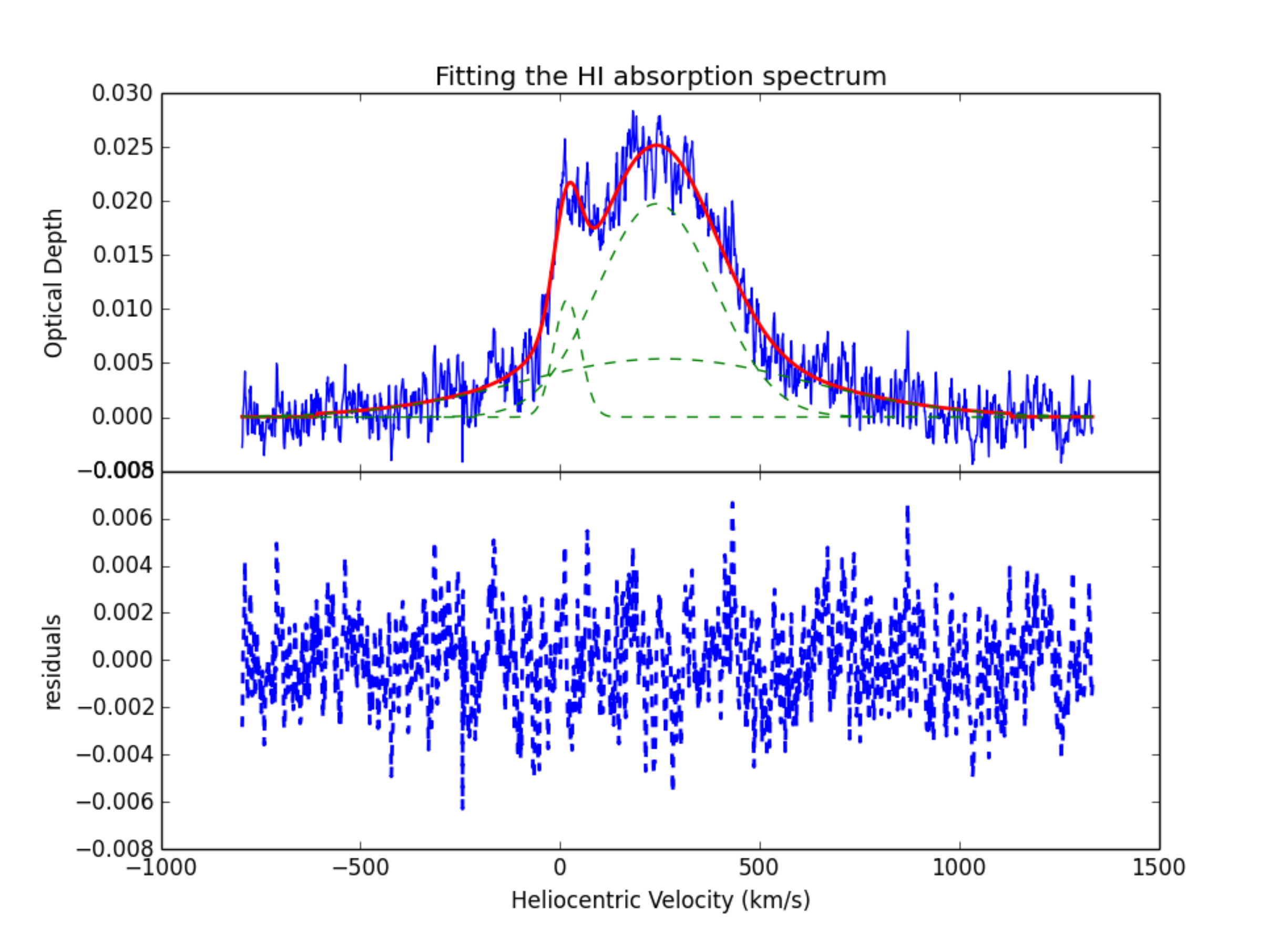}
\caption{\label{fig2} (Top) The optical depth of HI absorption for 4C+15.05; red line: summed Gaussian fits; green lines: individual Gaussian components; (Bottom) The difference between the data and the red line, demonstrating the goodness of fit. The $x$-axis is the recession velocity relative to the \cite{Stickel+96} redshift of $z=0.833$.}
\end{figure}

\section{Discussion}

\begin{deluxetable*}{cccccc}
\tabletypesize{\footnotesize}
\centering
\tablecaption{Multiple-Gaussian fits to the HI absorption spectrum of 4C+15.05}
\label{table2}
\tablehead{
\colhead{ID no} & \colhead{$v_{hel}$} & \colhead{z}  & \colhead{FWHM} & \colhead{Frac. abs.} & \colhead{$N$(HI)}
\\
\colhead{} & \colhead{km s$^{-1}$} & \colhead{} & \colhead{km s$^{-1}$} & \colhead{}  & \colhead{10$^{20}$cm$^{-2}$}
}

\startdata
1 & 249745(0.82)& 0.83306  & 78.7(1.90) & 0.0109 &  1.65(0.04) \\
2 & 249968(1.74)& 0.83380 & 345.9(6.61) & 0.0197 &  13.4(0.25) \\
3 & 249988(32.9)& 0.83387 & 848.1(128.2) & 0.00540 & 8.83(1.3) \\
\enddata
\end{deluxetable*}

\subsection{The Significance of a High N(HI)}
With a total neutral hydrogen column density of $N$(HI)$ =2.39 \times 10^{21}$ cm$^{-2}$, 4C+15.05 can  be considered a high HI column density source, on par with the deep absorption of J1148+5924 reported by ~\cite{Gupta+06}. Previous work \citep[such as][and references therein]{Pihlstrom03,Gupta+06} has suggested an anticorrelation between $N$(HI) and the projected or linear source size, which ~\cite{Curran13b} attribute to geometric effects. Some studies \citep[e.g.][] {Gupta+Saikia06,Kapahi+Saikia82} also find a correlation between the prominence of the core in a radio source and high $N$(HI) measurements. High resolution VLBI imaging at 5 GHz \citep{Linford+12} shows strong evidence for a central core component to 4C+15.05, on the order of 10 milliarcsec. At a redshift of $z=0.833$, this yields a projected linear size of $\sim$ 77 pc (assuming the following cosmological parameters: $H_{o}=69.6$ km s$^{-1}$ Mpc$^{-1}$, $\Omega_{M}=0.286$, and $\Omega_{vac}=0.714$). They also report a 5-GHz core flux density of 1.5 Jy, indicating (conservatively) that $\sim$ 60\% of the total 5-GHz flux density for the whole galaxy comes from a region on the order of a few milliarcsec. Between this higher core fraction, the small angular size of the core, and the high HI column density, 4C+15.05 is consistent with both the $N$(HI) vs core fraction and the $N$(HI) vs projected linear size correlations mentioned above. 

Overall, HI absorption has rarely been detected in high redshift ($z>1$) radio sources \citep[see, e.g.][]{KC03,Curran08}. Some studies \citep{CW12,Curran13a,Curran13c,Allison+12,Aditya+16} suggest what ~\cite{Curran+17} call a ``critical UV luminosity,'' $L_{UV_{crit}} = 10^{23}$ W Hz$^{-1}$, above which the neutral gas in a host galaxy is fully ionized. The theory is then that the lack of HI absorption in high-z samples stems from a selection effect, i.e. high-$z$ sample populations are looking further away, so they necessarily contain intrinsically high luminosity objects, particularly UV-luminous objects which would be capable of fully ionizing their neutral hydrogen. With the confirmation of $z=0.8336$, 4C+15.05 is a relatively high redshift source after all, and highly luminous as a blazar, so it is a bit of a curiosity in that it features HI absorption.

While 4C+15.05 has not been observed in the UV, we can estimate a $L_{UV}$ from publicly available data following the methods described in Appendix B of \cite{Curran08}. This requires first extrapolating to UV flux from observed photometry in two bands by fitting a power law. \cite{Curran08} in particular base their work predominantly on $B$ and $R$ band photometry, extrapolating to the flux at $\lambda_{U} = 1216$ \AA. At the redshift $z=0.8336$, this corresponds to $\lambda_{U}(1+z) =2229$ \AA, making $B$ or $V$ band photometry particularly important if we hope to make an accurate extrapolation. We use the \cite{Torrealba2012} $B=22.4$ mag, and a typical value from Vizier of $R=21.0$ mag, which for their respective band centers (440 nm and 680 nm) yield flux densities of $F_{B} = 0.00453$ mJy and $F_{R} = 0.0117$ mJy. At $z=0.8336$ (and assuming H$_{o} = 69.6$ km s$^{-1}$ Mpc$^{-1}$) this gives $L_{\lambda_{U}} \sim 10^{21}$ W $Hz^{-1}$, which is appropriately below the critical level at which the host galaxy would be expected to have its neutral gas fully ionized. However, since this blazar is known to vary in both the IR and optical, this prediction must be treated cautiously.


Without physically resolving the location of the neutral hydrogen detected by our observations, we cannot speak to whether these clouds are distinct components in the circumnuclear region, or part of a larger surrounding photodissociation region in which they are embedded, such as the dusty torus, or even part of a more global HI distribution within the host galaxy. Nevertheless, the three-component fit affirms the complex nature of the gas associated with the AGN. Interestingly (although the low level of accuracy of the optical redshift of the galaxy prevents us from saying anything definite), the three components do not appear to be partaking in larger scale movement, despite a blazar's potential for translating its energy into kinematic motion of the surrounding gas. The average redshift is distinct from any individual component, but two of the components (ID numbers 2 and 3) are centered at very similar central velocities, and the third is more significantly separated. This might (speculatively) be explained if the characteristic redshift of the whole host galaxy is better represented by those two components. If this is the case, then the first component (ID number 1) that produces the narrower `bump' on the side of the spectrum may perhaps be a signature of a blueshifted outflow driven towards us. High-resolution follow-up would be extremely valuable in order to distinguish the spatial characteristics of these kinematic features.

Classically, 4C+15.05 has been classified as a flat-spectrum source due to the flatness of its 300-3000 MHz continuum spectrum discussed in the introduction. Interestingly, ~\cite{An+16} suggest that 4C+15.05 may be considered a compact steep-spectrum (CSS) source instead, due to its compact size and a possible spectral flattening or even turnover at 200 MHz. Their spectral analysis is made based on reported values from the NASA Extragalactic Database, and as they point out, the different epochs at which the data were collected makes this source worthy of further study to confirm the turnover at low frequencies. The compactness of the source depends on high-resolution radio imaging found in \cite{An+16} and \cite{Wang+03}. With the accuracy of the redshift reported in this present work, the projected linear separation of 250 milliarcsec between the NE and SW ends of the 5-GHz MERLIN image corresponds to 1.94 kpc rather than the 1.3 kpc derived in \cite{An+16}. This is still well within the scale of the host galaxy ($<15$ kpc) and so does not preclude the identification of 4C+15.05 as a CSS. The dense ISM environment of a CSS, which is responsible for reining in the emission of the AGN, is an appropriate environment for high density neutral gas such as is reported in this paper to be found, even in proximity to the powerful outflows of a blazar.

\subsection{Concerning the Redshift Discrepancy}

Two of the previous works that report a redshift for 4C+15.05 present the spectra upon which the estimate is based. The near-IR/optical spectroscopy of \cite{Stickel+96} was taken with the Red Channel spectrograph on the Multiple Mirror Telescope and covered 4200 to 8299 \AA. This reveals narrow emission lines of [O II] 3729 and [Ne III] 3870, redshifted by $z=0.833$ and ``of moderate excitation.'' The resolution achieved was 18 \AA . The near-IR/optical spectroscopy of \cite{Perlman98} spans 4000-10000 \AA, and was taken with the ESO 3.6m telescope, with an unspecified (but ``improved'') spectral resolution. They observed the same lines that appear in the Stickel et al. spectra, but identified them as H-$\beta$ and [O III]. Additionally, in their higher resolution spectra they propose identification of H-$\alpha$ and H-$\eta$ as well as a Mg II line near 3800 \AA. With this interpretation, the spectra yields $z=0.405$. This conclusion is dependent on the identification of the H-$\alpha$ line, the primary difference between the Perlman et al and the Stickel et al results. While the Perlman et al work claims identification of additional lines, the signal-to-noise ratio makes them much less reliable indicators of redshift. Indeed, it is difficult to even distinguish the H-$\eta$ line from the continuum. Unfortunately, the line that Perlman et al. label as H-$\alpha$ is also in a noisier part of the continuum -- the sudden jaggedness of the spectrum at the red end suggests, perhaps, an inadequately subtracted sky line. Alternately, reconsidering the data from \cite{Perlman98} to be at the redshift of $z=0.833$, the proposed H-$\alpha$ line at about 9180 \AA\ can be interpreted as an [O III] doublet of 5088.927 and 5088.962 \AA.

\section{Conclusions}
Neutral hydrogen is a unique indicator in radio astronomy, distinct as it is from most other atomic and molecular lines. For a redshift of either $z=1.202$ or $z=0.405$, there is no other possible identification of the absorption observed in the spectrum of 4C+15.05 at $\sim$ 774.4 MHz. The correspondence of this absorption with neutral hydrogen shifted by the precise amount indicated by the \cite{Stickel+96} reported redshift of $z=0.833$ provides unambiguous evidence for establishing the redshift of 4C+15.05. Moreover, with the high resolution and sensitivity of Arecibo Observatory's 305 m telescope, we are able to improve the accuracy of this measurement by an order of magnitude, as well as reveal a complex kinematic structure of HI absorbers within the blazar. The presence of a high density component of neutral gas is additional evidence that may support the identification of 4C+15.05 as a compact steep spectrum source. 

\section{Acknowledgements}

This research has made use of the NASA/IPAC Extragalactic Database (NED) which is operated by the Jet Propulsion Laboratory, California Institute of Technology, under contract with the National Aeronautics and Space Administration. The team graciously thanks the National Science Foundation's Research Experience for Undergraduates program for funding work on the preliminary version of the 700-800 MHz receiver. The Arecibo Observatory is operated by SRI International under a cooperative agreement with the National Science Foundation (AST-1100968), and in alliance with Ana G. M\'endez-Universidad Metropolitana, and the Universities Space Research Association.

\bibliographystyle{aasjournal.bst}
\bibliography{redshift_4c1505_v8.bib}

\begin{thebibliography}{}
\expandafter\ifx\csname natexlab\endcsname\relax\def\natexlab#1{#1}\fi
\providecommand{\url}[1]{\href{#1}{#1}}

\bibitem[{{Aditya} {et~al.}(2016){Aditya}, {Kanekar}, \&
  {Kurapati}}]{Aditya+16}
{Aditya}, J.~N.~H.~S., {Kanekar}, N., \& {Kurapati}, S. 2016, \mnras, 455, 4000

\bibitem[{{Allison} {et~al.}(2012){Allison}, {Curran}, {Emonts}, {Ger{\'e}b},
  {Mahony}, {Reeves}, {Sadler}, {Tanna}, {Whiting}, \& {Zwaan}}]{Allison+12}
{Allison}, J.~R., {Curran}, S.~J., {Emonts}, B.~H.~C., {et~al.} 2012, \mnras,
  423, 2601

\bibitem[{{An} {et~al.}(2016){An}, {Cui}, {Baan}, {Wang}, \& {Mohan}}]{An+16}
{An}, T., {Cui}, Y.-Z., {Baan}, W.~A., {Wang}, W.-H., \& {Mohan}, P. 2016,
  \apj, 826, 190

\bibitem[{{Bennett} {et~al.}(1984){Bennett}, {Lawrence}, \&
  {Burke}}]{Bennett84}
{Bennett}, C.~L., {Lawrence}, C.~R., \& {Burke}, B.~F. 1984, \apjs, 54, 211

\bibitem[{{Condon} \& {Dressel}(1973)}]{Condon+1973}
{Condon}, J.~J., \& {Dressel}, L.~L. 1973, \aplett, 15, 203

\bibitem[{{Condon} \& {Jauncey}(1974{\natexlab{a}})}]{Condon+1974a}
{Condon}, J.~J., \& {Jauncey}, D.~L. 1974{\natexlab{a}}, \aj, 79, 437

\bibitem[{{Condon} \& {Jauncey}(1974{\natexlab{b}})}]{Condon74b}
---. 1974{\natexlab{b}}, \aj, 79, 1220

\bibitem[{{Curran} {et~al.}(2013{\natexlab{a}}){Curran}, {Allison}, {Glowacki},
  {Whiting}, \& {Sadler}}]{Curran13a}
{Curran}, S.~J., {Allison}, J.~R., {Glowacki}, M., {Whiting}, M.~T., \&
  {Sadler}, E.~M. 2013{\natexlab{a}}, \mnras, 431, 3408

\bibitem[{{Curran} {et~al.}(2017){Curran}, {Hunstead}, {Johnston}, {Whiting},
  {Sadler}, {Allison}, \& {Bignell}}]{Curran+17}
{Curran}, S.~J., {Hunstead}, R.~W., {Johnston}, H.~M., {et~al.} 2017, \mnras,
  470, 4600

\bibitem[{{Curran} \& {Whiting}(2012)}]{CW12}
{Curran}, S.~J., \& {Whiting}, M.~T. 2012, \apj, 759, 117

\bibitem[{{Curran} \& {Whiting}(2013)}]{Curran13b}
{Curran}, S.~J., \& {Whiting}, M.~T. 2013, in IAU Symposium, Vol. 292,
  Molecular Gas, Dust, and Star Formation in Galaxies, ed. T.~{Wong} \&
  J.~{Ott}, 243--243

\bibitem[{{Curran} {et~al.}(2013{\natexlab{b}}){Curran}, {Whiting}, {Sadler},
  \& {Bignell}}]{Curran13c}
{Curran}, S.~J., {Whiting}, M.~T., {Sadler}, E.~M., \& {Bignell}, C.
  2013{\natexlab{b}}, \mnras, 428, 2053

\bibitem[{{Curran} {et~al.}(2008){Curran}, {Whiting}, {Wiklind}, {Webb},
  {Murphy}, \& {Purcell}}]{Curran08}
{Curran}, S.~J., {Whiting}, M.~T., {Wiklind}, T., {et~al.} 2008, \mnras, 391,
  765

\bibitem[{{Dermer} \& {Schlickeiser}(1992)}]{Dermer+92}
{Dermer}, C.~D., \& {Schlickeiser}, R. 1992, Science, 257, 1642

\bibitem[{{Fan} {et~al.}(2008){Fan}, {Yuan}, {Liu}, {Hua}, {Romero}, {Zhang},
  {Su}, {Gupta}, {Liu}, {Huang}, {Guo}, {Zhang}, {Wang}, {Zhang}, \&
  {Tao}}]{Fan+2008}
{Fan}, J.-H., {Yuan}, Y.-H., {Liu}, Y., {et~al.} 2008, \pasj, 60, 707

\bibitem[{{Ghosh} \& {Rao}(1992)}]{Ghosh+Rao92}
{Ghosh}, T., \& {Rao}, A.~P. 1992, \aap, 264, 203

\bibitem[{{Ghosh} \& {Salter}(2002)}]{Ghosh+Salter02}
{Ghosh}, T., \& {Salter}, C. 2002, in Astronomical Society of the Pacific
  Conference Series, Vol. 278, Single-Dish Radio Astronomy: Techniques and
  Applications, ed. S.~{Stanimirovic}, D.~{Altschuler}, P.~{Goldsmith}, \&
  C.~{Salter}, 521--524

\bibitem[{{Giovanoni} \& {Kazanas}(1990)}]{Giovanoni+Kazanas90}
{Giovanoni}, P.~M., \& {Kazanas}, D. 1990, \nat, 345, 319

\bibitem[{{Gower} {et~al.}(1967){Gower}, {Scott}, \& {Wills}}]{Gower67}
{Gower}, J.~F.~R., {Scott}, P.~F., \& {Wills}, D. 1967, \memras, 71, 49

\bibitem[{{Gupta} \& {Saikia}(2006)}]{Gupta+Saikia06}
{Gupta}, N., \& {Saikia}, D.~J. 2006, \mnras, 370, 738

\bibitem[{{Gupta} {et~al.}(2006){Gupta}, {Salter}, {Saikia}, {Ghosh}, \&
  {Jeyakumar}}]{Gupta+06}
{Gupta}, N., {Salter}, C.~J., {Saikia}, D.~J., {Ghosh}, T., \& {Jeyakumar}, S.
  2006, \mnras, 373, 972

\bibitem[{{Impey} \& {Tapia}(1990)}]{Impey+Tapia90}
{Impey}, C.~D., \& {Tapia}, S. 1990, \apj, 354, 124

\bibitem[{{Kanekar} \& {Chengalur}(2003)}]{KC03}
{Kanekar}, N., \& {Chengalur}, J.~N. 2003, \aap, 399, 857

\bibitem[{{Kapahi} \& {Saikia}(1982)}]{Kapahi+Saikia82}
{Kapahi}, V.~K., \& {Saikia}, D.~J. 1982, Journal of Astrophysics and
  Astronomy, 3, 465

\bibitem[{{Kellermann} \& {Pauliny-Toth}(1969)}]{Kellermann+PT69}
{Kellermann}, K.~I., \& {Pauliny-Toth}, I.~I.~K. 1969, \apjl, 155, L71

\bibitem[{{Kellermann} {et~al.}(1971){Kellermann}, {Jauncey}, {Cohen},
  {Shaffer}, {Clark}, {Broderick}, {R{\"o}nn{\"a}ng}, {Rydbeck}, {Matveyenko},
  {Moiseyev}, {Vitkevitch}, {Cooper}, \& {Batchelor}}]{Kellermann+71}
{Kellermann}, K.~I., {Jauncey}, D.~L., {Cohen}, M.~H., {et~al.} 1971, \apj,
  169, 1

\bibitem[{{Linford} {et~al.}(2012){Linford}, {Taylor}, {Romani}, {Helmboldt},
  {Readhead}, {Reeves}, \& {Richards}}]{Linford+12}
{Linford}, J.~D., {Taylor}, G.~B., {Romani}, R.~W., {et~al.} 2012, \apj, 744,
  177

\bibitem[{{Lister} {et~al.}(2013){Lister}, {Aller}, {Aller}, {Homan},
  {Kellermann}, {Kovalev}, {Pushkarev}, {Richards}, {Ros}, \&
  {Savolainen}}]{Lister+13}
{Lister}, M.~L., {Aller}, M.~F., {Aller}, H.~D., {et~al.} 2013, \aj, 146, 120

\bibitem[{{Marscher}(1977)}]{Marscher+1977}
{Marscher}, A.~P. 1977, \apj, 216, 244

\bibitem[{{Owen} {et~al.}(1980){Owen}, {Spangler}, \& {Cotton}}]{Owen+1980}
{Owen}, F.~N., {Spangler}, S.~R., \& {Cotton}, W.~D. 1980, \aj, 85, 351

\bibitem[{{Pauliny-Toth} {et~al.}(1966){Pauliny-Toth}, {Wade}, \&
  {Heeschen}}]{PT66}
{Pauliny-Toth}, I.~I.~K., {Wade}, C.~M., \& {Heeschen}, D.~S. 1966, \apjs, 13,
  65

\bibitem[{{Perlman} {et~al.}(1998){Perlman}, {Padovani}, {Giommi}, {Sambruna},
  {Jones}, {Tzioumis}, \& {Reynolds}}]{Perlman98}
{Perlman}, E.~S., {Padovani}, P., {Giommi}, P., {et~al.} 1998, \aj, 115, 1253

\bibitem[{{Pihlstr{\"o}m} {et~al.}(2003){Pihlstr{\"o}m}, {Conway}, \&
  {Vermeulen}}]{Pihlstrom03}
{Pihlstr{\"o}m}, Y.~M., {Conway}, J.~E., \& {Vermeulen}, R.~C. 2003, \aap, 404,
  871

\bibitem[{{Stickel} {et~al.}(1996){Stickel}, {Rieke}, {Kuehr}, \&
  {Rieke}}]{Stickel+96}
{Stickel}, M., {Rieke}, G.~H., {Kuehr}, H., \& {Rieke}, M.~J. 1996, \apj, 468,
  556

\bibitem[{{Ter{\"a}sranta}(2000)}]{Terasranta+2000}
{Ter{\"a}sranta}, H. 2000, in American Institute of Physics Conference Series,
  Vol. 510, American Institute of Physics Conference Series, ed. M.~L.
  {McConnell} \& J.~M. {Ryan}, 352--356

\bibitem[{{Ter{\"a}sranta} {et~al.}(2005){Ter{\"a}sranta}, {Wiren}, {Koivisto},
  {Saarinen}, \& {Hovatta}}]{Terasranta+2005}
{Ter{\"a}sranta}, H., {Wiren}, S., {Koivisto}, P., {Saarinen}, V., \&
  {Hovatta}, T. 2005, \aap, 440, 409

\bibitem[{{Ter{\"a}sranta} {et~al.}(1992){Ter{\"a}sranta}, {Tornikoski},
  {Valtaoja}, {Urpo}, {Nesterov}, {Lainela}, {Kotilainen}, {Wiren}, {Laine},
  {Nilsson}, \& {Valtonen}}]{Terasranta+1992}
{Ter{\"a}sranta}, H., {Tornikoski}, M., {Valtaoja}, E., {et~al.} 1992, \aaps,
  94, 121

\bibitem[{{Ter{\"a}sranta} {et~al.}(1998){Ter{\"a}sranta}, {Tornikoski},
  {Mujunen}, {Karlamaa}, {Valtonen}, {Henelius}, {Urpo}, {Lainela}, {Pursimo},
  {Nilsson}, {Wiren}, {Laehteenmaeki}, {Korpi}, {Rekola}, {Heinaemaeki},
  {Hanski}, {Nurmi}, {Kokkonen}, {Keinaenen}, {Joutsamo}, {Oksanen},
  {Pietilae}, {Valtaoja}, {Valtonen}, \& {Koenoenen}}]{Terasranta+98}
{Ter{\"a}sranta}, H., {Tornikoski}, M., {Mujunen}, A., {et~al.} 1998, \aaps,
  132, 305

\bibitem[{{Torrealba} {et~al.}(2012){Torrealba}, {Chavushyan},
  {Cruz-Gonz{\'a}lez}, {Arshakian}, {Bertone}, \&
  {Rosa-Gonz{\'a}lez}}]{Torrealba2012}
{Torrealba}, J., {Chavushyan}, V., {Cruz-Gonz{\'a}lez}, I., {et~al.} 2012,
  \rmxaa, 48, 9

\bibitem[{{Wang} {et~al.}(2003){Wang}, {Hong}, {Jiang}, \& {An}}]{Wang+03}
{Wang}, W.~H., {Hong}, X.~Y., {Jiang}, D.~R., \& {An}, T. 2003, Acta
  Astronomica Sinica, 44, 299

\end{thebibliography}

\end{document}